\documentclass[12pt]{article}
\usepackage{amsmath}

\newcommand{\be}{\begin{equation}}
\newcommand{\ee}{\end{equation}}
\newcommand{\bea}{\begin{eqnarray}}
\newcommand{\eea}{\end{eqnarray}}
\newcommand{\bdm}{\begin{displaymath}}
\newcommand{\edm}{\end{displaymath}}

\def\T{{\widetilde T}}

\begin{document}
\title{\large \bf Apparent Superluminal Behavior\\
in Wave Propagation}
\author{A. D. Jackson\thanks{The Niels Bohr Institute, University of
Copenhagen, Blegdamsvej 17, 2100 Copenhagen,Denmark}, A.
Lande\thanks{Institute for Theoretical Physics, University of
Groningen, Nijenborgh 4, 9747AG Groningen, The Netherlands}, and B.
Lautrup$^*$}

\maketitle

\begin{abstract}
The apparent superluminal propagation of electromagnetic signals
seen in recent experiments is shown to be the result of simple and
robust properties of relativistic field equations. Although the wave
front of a signal passing through a classically forbidden region can
never move faster than light, an attenuated replica of the signal is
reproduced ``instantaneously'' on the other side of the barrier. The
reconstructed signal, causally connected to the forerunner rather
than the bulk of the input signal, appears to move through the
barrier faster than light.
\end{abstract}

\parskip=3mm
\parindent=0pt

Recent experimental reports of the propagation of electromagnetic
signals with velocities larger than $c$ in dispersive
media~\cite{DispersionExperiments}, in wave
guides~\cite{WaveGuideExperiments}, and in electronic
circuits~\cite{Mitchell} have once again focused attention on a
subject of long-standing interest. The immediate and wide-spread
hostility which such reports seem to generate among theorists is an
understandable consequence of a firm belief in the consistency of
electromagnetism and relativity.  In order to dispel such concerns
at the outset, we distinguish between ``true'' and ``apparent''
superluminal phenomena.  Consider a right-moving pulse with a wave
front located at $x_0$ at $t = 0$. A true superluminal phenomenon
would permit the observation of some signal at positions $x > x_0 +
ct$.  True superluminality is not predicted by either Maxwell's
equations or by the manifestly covariant Klein-Gordon equation which
we shall consider here. Indeed, most recent experimental papers are
careful to emphasize their consistency with Maxwell's equations and,
hence, do not claim the observation of true superluminal effects.
Rather, these experiments have demonstrated the existence of
``apparent'' superluminal phenomena taking place well behind the
light front.  In the case of microwave experiments, observations
generically involve pulses which seem to traverse a ``classically
forbidden'' region instantaneously.  While these results illuminate
an interesting and potentially useful effect, such transmission
occurs well behind the light front and does not challenge the wisdom
and authority of Maxwell and Einstein.  As we shall see below,
apparent superluminality is extremely general and to be expected.

Papers by Sommerfeld and Brillouin~\cite{ref2} represent some of the
earliest and most beautiful investigations of the question of
superluminality. Their concern was with unbounded, dispersive media.
There were at the time abundant examples of anomalous dispersion,
i.e.\ substances for which phase and group velocities were both
larger than $c$. Since the group velocity was then believed to be
identical to the velocity of energy propagation, Sommerfeld and
Brillouin understandably found the question of superluminal
propagation to be of importance.  Their strategy was to write the
requisite propagator using Laplace transforms and a suitable
analytic form for the phase velocity, $v ( \omega ) = \omega / k$.
The fact that the singularities of $v ( \omega )$ were restricted to
the lower half $\omega$ plane was then sufficient to prove that the
signal was necessarily zero ahead of the light front and that the
light front always moves with a velocity of $c$.  While the efforts
of Sommerfeld and Brillouin would seem to have settled the issue of
true superluminal propagation definitively, the situation is
somewhat more subtle.  Their work conclusively demonstrated that
Maxwell's equations preclude superluminal propagation for media with
a causal form of $v ( \omega )$.  It did not, however, extend to a
proof that the singularities of $v ( \omega )$  must lie in the lower
half plane. In simple electron resonance models of dielectrics, such
behavior follows from the absorptive nature of the
material~\cite{Jackson}.  We are not, however, aware of any
completely general proof of material causality.

The present paper addresses the current issue of apparent
superluminality. In order to avoid the difficult issue of modeling
$v( \omega )$, we will restrict our attention to propagation in
two-dimensional wave guides with constrictions. For slow variations
in the shape of the constriction, Maxwell's equations reduce to a
one-dimensional Klein-Gordon equation, in which the non-uniformities
can be modeled through a suitable potential.  A side benefit of this
replacement is that the strict impossibility of true superluminal
propagation is easily demonstrated.  We then consider the
propagation of a wave form with a precise front initially to the
left of a potential barrier located in the interval $0 \le x \le b$.
The barrier is presumed to be high relative to the dominant wave
numbers contained in the pulse but is otherwise arbitrary.  Our
results for the incoming and transmitted waves can then be expressed
simply: The incoming wave moves with a uniform velocity of $c=1$.
The transmitted wave $\psi(x,t)$ (for $x > b$) is  attenuated as an
obvious consequence of barrier penetration, and its amplitude is
proportional to the derivative of the initial pulse evaluated at the
point $(x - ct -b)$.  The additional displacement, $b$, suggests
instantaneous transmission of the pulse through the barrier and is
the source of the apparent superluminality observed empirically. The
fact that the transmitted pulse is an image of the derivative of the
original pulse and not the pulse itself is an elementary consequence
of the fact that  transmission amplitudes generally vanish in the
limit $\omega\to 0$.  When the signal is a low-frequency modulation
of a carrier wave, as is the case in many experimental
investigations, the envelopes of the incident and transmitted waves
are identical.

Some of the topics treated here have been discussed elsewhere both
analytically and numerically \cite{Emmich,Mitchell}. Our intention
is to emphasize the generality and extreme simplicity of this
phenomenon.

\paragraph{The Model:}  We consider a scalar wave, $\Psi$, moving in
a two-dimensional wave guide according to the Klein-Gordon equation
\begin{equation}
\nabla^2 \Psi = \frac{\partial^2 \Psi}{\partial t^2} \ . \label{1}
\end{equation}
The wave guide is infinite in the $x$-direction and extends from $0
\le z \le h(x)$ in the $z$-direction.  We assume that $\Psi$
vanishes at the transverse bounding surfaces.  If $h$ is a slowly
varying function of $x$, we can approximate $\Psi$ as the product
$\psi (x,t) \sin{(\pi z / h)}$. Neglecting derivatives of $h$,
eqn.\,(\ref{1}) reduces to a one-dimensional equation:
\begin{equation}
- \frac{\partial^2 \psi}{\partial x^2} + V(x) \psi = -
\frac{\partial^2 \psi}{\partial t^2} \ . \label{2}
\end{equation}
The effective potential is determined by the width of the wave guide
so that $V(x) = \pi^2 /h(x)^2$ for the lowest transverse mode.  For
simplicity, we assume that $h(x)$ is  large except in the vicinity
of the constriction, so that $V(x)$ can be modeled as non-zero only
in the region $0 \le x \le b$.

We seek solutions to eqn.\,(\ref{2}) which describe the motion of an
initial pulse, $\psi (x,0) = f(x)$, which has arbitrary shape but
satisfies the following two conditions.  First, the pulse has a
well-defined wave front, $x_0$, initially to the left of the
barrier, $V(x)$, i.e.,  $\psi (x , 0) \ne 0$ only when $x \le x_0 <
0$.  Second, at $t=0$ the pulse moves uniformly to the right with a
velocity of $1$, so that $\partial \psi /
\partial t = - \partial \psi / \partial x$ at $t = 0$.
For any given potential, this problem can be solved with the aid of
the corresponding Green's function.

This model has been considered in detail in ref.~\cite{Deutch&Low},
where it was shown in generality that the transmitted wave is given
by
\begin{equation}
\psi(x,t)=\int_{-\infty}^{x_0} \T(x-t-x') f(x')\,dx' \ , \label{3}
\end{equation}
where $T(x)$ is the retarded transmission kernel, which may be
expressed in terms of its Fourier transform $T(\omega)$
\begin{equation}
\T(u)=\int_{-\infty}^\infty T(\omega)e^{i\omega
u}\,\frac{d\omega}{2\pi}~. \label{4}
\end{equation}
The physical interpretation of $T( \omega )$ as a transmission
amplitude is elementary:  An incoming plane wave $\exp{(i \omega
x)}$, incident on the potential barrier from the left, leads to a
transmitted wave $T( \omega ) \, \exp{(i \omega x)}$.  Since
$|T(\omega)|\to1$ for $|\omega|\to\infty$, the integration contour
in eqn.\,(\ref{4}) can be closed in the upper half $\omega$-plane for
$x>0$.  If $T ( \omega )$ is free of singularities in the upper half
plane, it follows that $\T(x-t-x')=0$ for $x>x'+t$ and thus that
$\psi (x,t)$ is strictly zero for $x > x_0+t$. Nothing precedes the
light front.

\paragraph{A Special Case:}

The authors of ref.~\cite{Deutch&Low} considered the special case
where $V(x)=m^2$ is a positive constant and found
\begin{equation}
T( \omega ) = \frac{4 \omega \kappa}{D} \, {\rm e}^{i (\kappa -
\omega )b} \label{5}
\end{equation}
with $\kappa= i \sqrt{m^2 - (\omega+i\epsilon)^2}$
and
\begin{equation}
 D =
(\omega + \kappa )^2 - (\omega - \kappa )^2 {\rm e}^{2 i \kappa b} \
. \label{7}
\end{equation}
The singularities of $T( \omega )$ are due to the zeros of $D$ in
the $\omega$-plane.  Given the form of $\kappa$, the zeros of $D$
are confined to the lower half plane, and $T( \omega )$ is indeed
analytic in the upper half plane.  As expected, this model precludes
genuine superluminal propagation.  The analytic properties of $T(
\omega )$ are, of course, dictated by those of $V( \omega )$.  The
general proof that any given potential will lead to such analyticity
is more challenging.\footnote{A general proof can be constructed
along the following lines.  Write the transmission amplitude as a
linear integral equation of the form $\psi= \varphi + \int \,
G_0\,V\,\psi$, where $G_0$ is a suitable free propagator.
Singularities in $\psi$ arise when singularities of the integrand
pinch the integration contour. Analyticity of $V( \omega )$ in the
upper half plane then ensures the desired analyticity properties of
$T ( \omega )$.}  All real, local, and bounded potentials which
vanish sufficiently rapidly as $x \to \infty$ are expected to
respect these analyticity conditions, and the absence of true
superluminality is thus to be expected for all physically sensible
choices of $V$.

\paragraph{Apparent Superluminal Behaviour:}
Confident that our Klein-Gordon model is free of genuine acausal
propagation, we turn  to apparent superluminal phenomena.  Consider
a strong barrier (with $mb >> 1$) and imagine that the initial wave
form, $\psi (x,0)$, is dominated by low frequency components for
which $|\omega| << m$.  In this case, the form of $\psi (x,t)$ is
both simple and intuitive. Specifically, we need only consider
$\omega\approx0$ for which $\kappa\approx im$. In this domain, the
transmission amplitude can be approximated as
\begin{equation}
 T( \omega ) \approx - \omega \frac{4i}{m} {\rm e}^{-i \omega b}
{\rm e}^{- m b} \ . \label{8}
\end{equation}
We shall see shortly that this form of the transmission amplitude is
quite general.  Using eqn.\,(\ref{8}), we see that eqn.\,(\ref{4})
reduces to
\begin{equation}
 \T(x-t-x')=-\frac4m e^{-mb}\left.\frac\partial{\partial
 u}\delta(u-b)\right|_{u=x-t-x'}
\label{9}
\end{equation}
Thus, we find that
\begin{equation}
 \psi (x ,t) \approx -\frac4m{\rm e}^{-bm} f'(x-t-b)\ .
\label{10}
\end{equation}
When a pulse dominated by low frequencies components impinges on a
strongly repulsive barrier, the transmitted wave is a strongly
attenuated replica of the derivative of the original pulse. The
transmitted pulse appears to traverse the region of the potential
barrier in zero time.  This is the apparent superluminal phenomenon
observed empirically.  It occurs well behind the light front of the
original signal and is not an indication of true superluminal
propagation. Rather, it is an interference phenomenon which is in no
sense acausal.

There is an evident inconsistency between the present assumption that
$\psi (x , 0)$ is dominated by low frequency components and our
initial assumption that the signal has a well-defined light front
(which necessarily implies the presence of high frequency
components). The consideration of signals which are the product of,
e.g., a gaussian pulse (with clear low-frequency dominance) and a
step function to impose the light front makes it clear that the
effects of this inconsistency can be made arbitrarily
small~\cite{Deutch&Low}.

\paragraph{Carrier Waves:} Experiments frequently involve a modulated carrier
wave,
\begin{equation}
f(x) = {\rm e}^{i \omega_0 x} \, F(x) \ , \label{11}
\end{equation}
where $F(x)$ provides a slowly-varying modulation of the carrier.
Inserting this into eqn.\,(\ref{10}), the transmitted signal becomes
\begin{equation}\label{12}
\psi(x,t)=-\frac4m e^{-bm}\left.\left(i\omega_0
F(u)+F'(u)\right)e^{i\omega_0 u}\right|_{u=x-t-b}
\end{equation}
Since the Fourier transform of $F(x)$ is presumed to have support
only for frequencies $|\omega|\ll\omega_0$, the second term may be
ignored. We conclude that the envelope of the pulse, $| \psi (x,t)|$,
is unaltered by the transmission. Again, the argument of the right
side of eqn.\,(\ref{12}) suggests that transmission of the envelope
through the barrier is instantaneous.

\paragraph{Generality of the Results:}
The various factors contributing to the approximate form,
eqn.\,(\ref{8}), for $T( \omega )$ are all of general origin.  The
factor $\exp{(-i \omega b)}$ represents the phase difference between
the free plane wave, \break $\exp{(i \omega x)}$, at the boundaries
of the region of non-zero potential.  It will always appear.
Similarly, the linear vanishing of the transmission amplitude as
$\omega \to 0$ is a feature common  to all potentials which do not
have a zero-energy bound state.\footnote{Consider scattering from an
arbitrary potential which is zero except in the interval $0 \le x
\le b$ with an interior solution $\varphi (x)$. Join this interior
solution to the right-moving plane wave, $\exp{i \omega (x-b)}$, at
$x=b$ with the usual requirement of continuity of the wave function
and its first derivative.  In the limit $\omega \to 0$, $\varphi (b)
= 1$ and $\varphi' (b) = 0$. Similarly, join the interior solution
to the linear combination $A \exp{(i \omega x)} + B \exp{(-i \omega
x)}$ at $x = 0$.  If $\varphi' (0) \ne 0$, the coefficients $A$ and
$B$ will diverge like $1/\omega$.  The transmission amplitude, $T(
\omega ) = e^{-i\omega b}/A$, will vanish linearly with $\omega$
unless $\varphi' (0)$ is zero.  The condition that $\varphi' (0) =
\varphi' (b) = 0$ is precisely the condition that the potential
should support a zero-energy bound state. }

The final barrier penetration factor is also familiar and is
expected whenever there is strong attenuation. Consider the
transmission amplitude for a strongly repulsive but otherwise
arbitrary potential using the WKB approximation.\footnote{We
consider a potential which is strongly repulsive for all $0 \le x
\le b$ and zero elsewhere.  Hence, it is appropriate to match the
plane wave solutions directly to the WKB wave function.} The
resulting transmission amplitude is readily calculated, and shows
that the factor $\exp(-mb)$ is replaced by
\begin{equation}
\exp\left[-\int_0^b \, \sqrt{V(x)} \, dx \right] \ . \label{15}
\end{equation}
A localized repulsive barrier of sufficient strength will transmit
an instantaneous image of the derivative of the incoming signal
according to eqn.\,(\ref{10}) independent of the details of both the
potential barrier and the pulse. Apparent superluminal behavior is a
robust phenomenon.

\paragraph{The Time Delay:}
The above results can also be expressed as a time delay of the pulse,
$\tau$, defined as the difference between the time actually required
for transmission across the barrier less the time required for a free
wave to travel the same distance.  In the case of the square barrier
and a low frequency pulse, $\tau = -b$.  Negative values of $\tau$
correspond to apparent superluminal propagation. For a modulated
carrier wave, eqn.\,(\ref{11}), one may expand $T ( \omega ) $ about
the carrier frequency, $\omega_0$, and obtain
\begin{equation}
T ( \omega ) \approx | T ( \omega_0 ) | {\rm e}^{i \Phi ( \omega_0
)} \ {\rm e}^{i ( \omega - \omega_0 ) \Phi'(\omega_0)} \ , \label{16}
\end{equation}
where $\Phi(\omega)$ is the phase of $T(\omega)$. The second
exponential factor gives rise to the time delay, $\tau ( \omega_0 )
= \Phi' ( \omega_0 )$, through the Fourier exponential in
eqn.\,(\ref{4}).  Familiar results from quantum mechanics for purely
repulsive potentials remind us that $\Phi ( \omega )$ is less than
$0$ for all $\omega$ and that $\Phi (0) = \Phi ( \infty ) =
0$.\footnote{This second result is a consequence of Levinson's
theorem which relates the asymptotic behavior of the phase shift to
the number of bound states supported by the potential, $n$, through
$\Phi (0) - \Phi ( \infty ) = n \pi$.  There are no bound states for
purely repulsive potentials.} The time delay is necessarily negative
for sufficiently small $\omega_0$ and apparent superluminal effects
can be observed for all repulsive potentials.  The time delay
changes sign for some value of $\omega_0$ comparable to the height
of the potential barrier, and it approaches zero from above as
$\omega_0$ tends to infinity. Apparent superluminality is a very
general phenomenon.

\paragraph{Conclusions:}  Using the model of a Klein-Gordon equation
with a potential, we have presented a simple description of the
apparent superluminal phenomena seen in wave guides:  Low frequency
waves seem to traverse such barriers in zero time.  Strongly
repulsive barriers always transmit an attenuated image of the
spatial derivative of the incident signal.  When the signal consists
of a modulated carrier wave, the envelope of this wave is
transmitted unaltered.  We have attempted to demonstrate that the
phenomena described here are both extremely general and
non-controversial.  Their experimental observation does not
challenge received wisdom and in no sense compromises our confidence
in general notions of causality.  It will be interesting to see
whether this interesting and general consequence of wave theory will
have practical applications.

\end{document}